\documentclass{article}

\usepackage{arxiv}

\usepackage[utf8]{inputenc} 
\usepackage[T1]{fontenc}    
\usepackage{hyperref}       
\usepackage{url}            
\usepackage{booktabs}       
\usepackage{amsfonts}       
\usepackage{nicefrac}       
\usepackage{microtype}      
\usepackage{lipsum}
\usepackage{graphicx}
\graphicspath{ {./images/} }
\usepackage[authoryear, round]{natbib} 
\usepackage{authblk} 

\title{PySEMTools: A library for post-processing hexahedral spectral element data}

\author[1]{Adalberto Perez} 
\author[2]{Siavash Toosi} 
\author[4]{Tim Felle Olsen} 
\author[3]{Stefano Markidis} 
\author[1,2]{Philipp Schlatter} 

\affil[1]{FLOW Dept.  Engineering Mechanics, KTH Royal Institute of Technology}
\affil[2]{Institute of Fluid Mechanics (LSTM), Friedrich--Alexander--Universität (FAU)}
\affil[3]{Division of Computational Science and Technology (CST), KTH Royal Institute of Technology}
\affil[4]{Department of Civil and Mechanical Engineering Solid Mechanics, Technical University of Denmark}

\begin{document}
\maketitle


\section{Summary}
PySEMTools is a Python-based library for post-processing simulation data produced with high-order hexahedral elements in the context of the spectral element method in computational fluid dynamics. It aims to minimize intermediate steps typically needed when analyzing large files. Specifically, the need to use separate codebases (like the solvers themselves) at post-processing. For this effect, we leverage the use of message passing interface (MPI) for distributed computing to perform typical data processing tasks such as spectrally accurate differentiation, integration, interpolation, and reduced order modeling, among others, on a spectral element mesh. All the functionalities are provided in self-contained Python code and do not depend on the use of a particular solver. We believe that PySEMTools provides tools to researchers to accelerate scientific discovery and reduce the entry requirements for the use of advanced methods in computational fluid dynamics.

\section{Statement of need}
The motion of fluids around objects is fundamental for many industrial and natural systems, from aerodynamics and cooling to the behavior of weather systems. Particularly relevant applications generally exist in the turbulent flow regime, where the fluid is subject to unsteady motions that are characterized by the interactions of eddies of multiple sizes and where increased levels of fluctuations and mixing exist.

A popular method to study these phenomena is to use computers to simulate their governing physics. The multi-scale characteristic of turbulence and the high Reynolds numbers (ratio between inertial and viscous forces) that are typically of interest require that the numerical grids are fine enough to capture the motion of the smallest eddies. While this has implied that the computational cost of simulations is high, the advent of graphics processing units (GPUs) has opened the doors to perform simulations that would not have been possible in the past. This increase in capability has made managing the data produced on a typical simulation campaign more challenging. Our work in PySEMTools aims to streamline the data management and post-processing of the results obtained from a particular numerical method often used to study turbulent flows while keeping high-order accuracy.

PySEMTools aims to help post-processing data from solvers that use the spectral element method (SEM) originally proposed by \cite{PATERA1984}, which is a high-order variant of the finite element method (FEM). In SEM, the computational domain is divided into a finite set of elements in which a Gauss-Lobatto-Legendre (GLL) grid of a given degree $N$ is embedded. Inside each element, the solution is expanded using polynomials of order $P = N - 1$, resulting in low dissipation and dispersive errors. 

Nek5000 \citep{nek5000-web-page}, written in Fortran 77, is a successful implementation of SEM that has been used for several studies on the field, such as simulations of vascular flows by \cite{fischer2006}, turbulent pipe flow by \cite{elkhoury2013}, flow around wings by \cite{mallor2024} and even nuclear applications as shown in the overview by \cite{merzari2020}. In general, the post-processing pipeline has been somewhat complicated, as when the data is needed in the SEM format, for example, to calculate derivatives of velocity fields, the solver itself has been used in a "post-processing" mode. This mode uses the solver and additional Fortran code that needs to be compiled to produce smaller files that can be used in Matlab or Python with e.g. PyMech \citep{pymech}, to perform signal processing, create plots, etc. NekRS \citep{fischer2021nekrs},  a GPU version of Nek5000, and Neko \citep{jansson2024, jansson2023}, a modern Fortran implementation of SEM have followed the same approach, motivating the necessity of our PySEMTools for the future.

The motivation behind using the solvers themselves with the data in its raw format is understandable, as these large files need to be processed in parallel due to their sheer size. Still, we believe that the process has become very cumbersome as multiple code bases need to be maintained for post-processing the data. With PySEMTools we have brought a solution to this, as we have included all the functionalities that are typically needed from the solvers while ensuring that the codes perform efficiently in parallel while also taking advantage of the rich library ecosystem present in Python.

\section{Features}

PySEMTools relies heavily on MPI for Python by \cite{mpi4py}, given that it has been designed from the beginning to work on distributed settings. For computations, we rely on NumPy \citep{numpy}. It has been extensively tested on data produced by Nek5000 and Neko but, as mentioned before, the implemented methods and routines are consistent with any SEM-like data structure with hexahedral elements. Among its most relevant features are the following:

\begin{itemize} 
    \item \textbf{Parallel IO}: A set of routines to perform distributed IO on Nek5000/Neko field files and directly keep the data in memory on NumPy arrays or PyMech data objects.
    \item \textbf{Parallel data interfaces}: A set of objects that aim to facilitate the transfer of messages among processors. Done to ease the use of MPI functions for more inexperienced users.
    \item \textbf{Calculus}:  Objects to calculate the derivation and integration matrices based on the geometry, which allows to perform calculus operations on the spectral element mesh.
    \item \textbf{Mesh connectivity and partitioning}: Objects to determine the connectivity based on the geometry and mesh repartitioning tools for tasks such as global summation, among others.
    \item \textbf{Interpolation}: Routines to perform high-order interpolation from an SEM mesh into any arbitrary query point. A crucial functionality when performing post-processing.
    \item \textbf{Reduced-order modeling}: Objects to perform parallel and streaming proper orthogonal decomposition (POD).
    \item \textbf{Data compression/streaming}: Through the use of ADIOS2 \citep{adios2}, a set of interfaces is available to perform data compression or to connect Python scripts to running simulations to perform in-situ data processing. 
    \item \textbf{Visualization}: Given that the data is available in Python, visualizations can be performed from readily available packages. 

\end{itemize}

We note that all of these functionalities are supported by examples in the software repository at \url{https://github.com/ExtremeFLOW/pySEMTools}.

\section{Acknowledgements}

This work is partially funded by the “Adaptive multi-tier intelligent data manager for Exascale (ADMIRE)” project, which is funded by the European Union's Horizon 2020 JTI-EuroHPC research and innovation program under grant Agreement number: 956748. Computations for testing were enabled by resources provided by the National Academic Infrastructure for Super­computing in Sweden (NAISS), partially funded by the Swedish Research Council through grant agreement no. 2018-05973.

\bibliographystyle{plainnat}  
\bibliography{pysemtools}  

\begin{thebibliography}{13}
\providecommand{\natexlab}[1]{#1}
\providecommand{\url}[1]{\texttt{#1}}
\expandafter\ifx\csname urlstyle\endcsname\relax
  \providecommand{\doi}[1]{doi: #1}\else
  \providecommand{\doi}{doi: \begingroup \urlstyle{rm}\Url}\fi

\bibitem[Dalcín et~al.(2005)Dalcín, Paz, and Storti]{mpi4py}
Lisandro Dalcín, Rodrigo Paz, and Mario Storti.
\newblock Mpi for python.
\newblock \emph{Journal of Parallel and Distributed Computing}, 65\penalty0 (9):\penalty0 1108--1115, 2005.
\newblock ISSN 0743-7315.
\newblock \doi{https://doi.org/10.1016/j.jpdc.2005.03.010}.

\bibitem[El~Khoury et~al.(2013)El~Khoury, Schlatter, Noorani, Fischer, Brethouwer, and Johansson]{elkhoury2013}
George~K. El~Khoury, Philipp Schlatter, Azad Noorani, Paul~F. Fischer, Geert Brethouwer, and Arne~V. Johansson.
\newblock Direct numerical simulation of turbulent pipe flow at moderately high reynolds numbers.
\newblock \emph{Flow, Turbulence and Combustion}, 91\penalty0 (3):\penalty0 475--495, Oct 2013.
\newblock ISSN 1573-1987.
\newblock \doi{10.1007/s10494-013-9482-8}.
\newblock URL \url{https://doi.org/10.1007/s10494-013-9482-8}.

\bibitem[Fischer et~al.(2006)Fischer, Loth, Lee, Smith, Tufo, and Bassiouny]{fischer2006}
Paul Fischer, Francis Loth, Sang-Wook Lee, David Smith, Henry Tufo, and Hisham Bassiouny.
\newblock - parallel simulation of high reynolds number vascular flows.
\newblock In Anil Deane, Akin Ecer, James McDonough, Nobuyuki Satofuka, Gunther Brenner, David~R. Emerson, Jacques Periaux, and Damien Tromeur-Dervout, editors, \emph{Parallel Computational Fluid Dynamics 2005}, pages 219--226. Elsevier, Amsterdam, 2006.
\newblock ISBN 978-0-444-52206-1.
\newblock \doi{https://doi.org/10.1016/B978-044452206-1/50026-4}.
\newblock URL \url{https://www.sciencedirect.com/science/article/pii/B9780444522061500264}.

\bibitem[Fischer et~al.(2008)Fischer, Lottes, and Kerkemeier]{nek5000-web-page}
Paul Fischer, James~W. Lottes, and Stefan~G. Kerkemeier.
\newblock {Nek5000} {w}eb page, 2008.
\newblock \url{http://nek5000.mcs.anl.gov}.

\bibitem[Fischer et~al.(2021)Fischer, Kerkemeier, Min, Lan, Phillips, Rathnayake, Merzari, Tomboulides, Karakus, Chalmers, et~al.]{fischer2021nekrs}
Paul Fischer, Stefan Kerkemeier, Misun Min, Yu-Hsiang Lan, Malachi Phillips, Thilina Rathnayake, Elia Merzari, Ananias Tomboulides, Ali Karakus, Noel Chalmers, et~al.
\newblock Nekrs, a gpu-accelerated spectral element navier-stokes solver.
\newblock \emph{arXiv preprint arXiv:2104.05829}, 2021.

\bibitem[Godoy et~al.(2020)Godoy, Podhorszki, Wang, Atkins, Eisenhauer, Gu, Davis, Choi, Germaschewski, Huck, et~al.]{adios2}
William~F Godoy, Norbert Podhorszki, Ruonan Wang, Chuck Atkins, Greg Eisenhauer, Junmin Gu, Philip Davis, Jong Choi, Kai Germaschewski, Kevin Huck, et~al.
\newblock Adios 2: The adaptable input output system. a framework for high-performance data management.
\newblock \emph{SoftwareX}, 12:\penalty0 100561, 2020.

\bibitem[Harris et~al.(2020)Harris, Millman, van~der Walt, Gommers, Virtanen, Cournapeau, Wieser, Taylor, Berg, Smith, Kern, Picus, Hoyer, van Kerkwijk, Brett, Haldane, del R{\'{i}}o, Wiebe, Peterson, G{\'{e}}rard-Marchant, Sheppard, Reddy, Weckesser, Abbasi, Gohlke, and Oliphant]{numpy}
Charles~R. Harris, K.~Jarrod Millman, St{\'{e}}fan~J. van~der Walt, Ralf Gommers, Pauli Virtanen, David Cournapeau, Eric Wieser, Julian Taylor, Sebastian Berg, Nathaniel~J. Smith, Robert Kern, Matti Picus, Stephan Hoyer, Marten~H. van Kerkwijk, Matthew Brett, Allan Haldane, Jaime~Fern{\'{a}}ndez del R{\'{i}}o, Mark Wiebe, Pearu Peterson, Pierre G{\'{e}}rard-Marchant, Kevin Sheppard, Tyler Reddy, Warren Weckesser, Hameer Abbasi, Christoph Gohlke, and Travis~E. Oliphant.
\newblock Array programming with {NumPy}.
\newblock \emph{Nature}, 585\penalty0 (7825):\penalty0 357--362, September 2020.
\newblock \doi{10.1038/s41586-020-2649-2}.
\newblock URL \url{https://doi.org/10.1038/s41586-020-2649-2}.

\bibitem[Jansson et~al.(2023)Jansson, Karp, Perez, Mukha, Ju, Liu, P\'{a}ll, Laure, Weinkauf, Schumacher, Schlatter, and Markidis]{jansson2023}
Niclas Jansson, Martin Karp, Adalberto Perez, Timofey Mukha, Yi~Ju, Jiahui Liu, Szil\'{a}rd P\'{a}ll, Erwin Laure, Tino Weinkauf, J\"{o}rg Schumacher, Philipp Schlatter, and Stefano Markidis.
\newblock Exploring the ultimate regime of turbulent rayleigh–b\'{e}nard convection through unprecedented spectral-element simulations.
\newblock In \emph{Proceedings of the International Conference for High Performance Computing, Networking, Storage and Analysis}, SC '23, New York, NY, USA, 2023. Association for Computing Machinery.
\newblock ISBN 9798400701092.
\newblock \doi{10.1145/3581784.3627039}.
\newblock URL \url{https://doi.org/10.1145/3581784.3627039}.

\bibitem[Jansson et~al.(2024)Jansson, Karp, Podobas, Markidis, and Schlatter]{jansson2024}
Niclas Jansson, Martin Karp, Artur Podobas, Stefano Markidis, and Philipp Schlatter.
\newblock Neko: A modern, portable, and scalable framework for high-fidelity computational fluid dynamics.
\newblock \emph{Computers \& Fluids}, 275:\penalty0 106243, 2024.
\newblock ISSN 0045-7930.
\newblock \doi{https://doi.org/10.1016/j.compfluid.2024.106243}.

\bibitem[Mallor et~al.(2024)Mallor, Vinuesa, Örlü, and Schlatter]{mallor2024}
Fermin Mallor, Ricardo Vinuesa, Ramis Örlü, and Philipp Schlatter.
\newblock High-fidelity simulations of the flow around a naca 4412 wing section at high angles of attack.
\newblock \emph{International Journal of Heat and Fluid Flow}, 110:\penalty0 109590, 2024.
\newblock ISSN 0142-727X.
\newblock \doi{https://doi.org/10.1016/j.ijheatfluidflow.2024.109590}.
\newblock URL \url{https://www.sciencedirect.com/science/article/pii/S0142727X24003151}.

\bibitem[Merzari et~al.(2020)Merzari, Fischer, Min, Kerkemeier, Obabko, Shaver, Yuan, Yu, Martinez, Brockmeyer, Fick, Busco, Yildiz, and Hassan]{merzari2020}
Elia Merzari, Paul Fischer, Misun Min, Stefan Kerkemeier, Aleksandr Obabko, Dillon Shaver, Haomin Yuan, Yiqi Yu, Javier Martinez, Landon Brockmeyer, Lambert Fick, Giacomo Busco, Alper Yildiz, and Yassin Hassan.
\newblock Toward exascale: Overview of large eddy simulations and direct numerical simulations of nuclear reactor flows with the spectral element method in nek5000.
\newblock \emph{Nuclear Technology}, 206\penalty0 (9):\penalty0 1308--1324, 2020.
\newblock \doi{10.1080/00295450.2020.1748557}.
\newblock URL \url{https://doi.org/10.1080/00295450.2020.1748557}.

\bibitem[Mohanan et~al.(2022)Mohanan, Chauvat, Kleine, Fabbiane, and Canton]{pymech}
Ashwin~Vishnu Mohanan, Guillaume Chauvat, Vitor~Gabriel Kleine, Nicolò Fabbiane, and Jacopo Canton.
\newblock Pymech: A python software suite for nek5000 and simson, 2022.

\bibitem[Patera(1984)]{PATERA1984}
Anthony~T Patera.
\newblock A spectral element method for fluid dynamics: Laminar flow in a channel expansion.
\newblock \emph{Journal of Computational Physics}, 54\penalty0 (3):\penalty0 468--488, 1984.
\newblock ISSN 0021-9991.
\newblock \doi{https://doi.org/10.1016/0021-9991(84)90128-1}.

\end{thebibliography}

\end{document}